\documentstyle[12pt]{article}
\begin{document}
\baselineskip 1.5pc
\def\ii{\'{\i}}
\def\bb{$\beta\beta_{2\nu}$}
\def\nd{$^{150}Nd$}
\def\sm{$^{150}Sm$}
\bigskip
\bigskip
\centerline{\bf\large Double beta decay in heavy deformed nuclei:}
\centerline{\bf\large what have we learned?}
\vskip 1cm
\centerline{Jorge G. Hirsch}
\centerline{\tenit Departamento de F\'{\i}sica, Centro de Investigaci\'on
y de Estudios Avanzados del IPN,}
\centerline{\tenit A. P. 14-740 M\'exico 07000 D.F.}
\bigskip
\centerline{\bf{Abstract:}}
\bigskip
The pseudo SU(3) approach is used to describe
low lying states and BE(2) intensities of rare
earth and actinide nuclei which are $\beta\beta$ decay candidates.
The $\beta\beta$ half lives of some of these nuclei to the ground and excited
states of the final ones are
evaluated for both the two and zero neutrino emitting modes.
The existence of selection rules which strongly restricts the decays is
discussed.  These restrictions represent a possible test of the model.
Up to now the  predictions are in
good agreement with the available experimental data.

\bigskip
\bigskip
\centerline{\bf{Resumen:}}
\bigskip
El formalismo pseudo SU(3) es usado para describir los estados de
menor energ\'{\i}a y sus intensidades BE(2) de n\'ucleos en la regi\'on de las
tierras raras y los actin\'{\i}dos que son candidatos a decaimientos
$\beta\beta$. Las vidas medias de esta desintegraci\'on  para algunos de
estos n\'ucleos al estado fundamental y estados excitados del n\'ucleo hijo
son evaluadas en los modos con y sin emisi\'on de neutrinos.
Se discute la existencia de reglas de selecci\'on que restringen fuertemente
los decaimientos y representan una posible prueba para el modelo.
Hasta el momento el acuerdo entre las predicciones y los datos experimentales
disponibles es bueno.


\vfill

\section{Introduction}
\bigskip

The detection of the neutrinoless double beta decay ($\beta\beta_{0\nu}$)
  would imply an indisputable evidence of physics beyond the
standard model and would be useful in order to select Grand
Unification Theories\cite{Ver86}.
Theoretical nuclear matrix elements are needed to convert
experimental half-life limits, which are available for many
$\beta\beta$-unstable isotopes\cite{Moe93}, into constrains for
particle physics parameters such as the effective Majorana mass of the
neutrino and the contribution of right-handed currents to the weak
interaction. The two neutrino mode of the double beta decay (\bb ) is
allowed as a second order process in the standard model. It has
been detected in nine nuclei\cite{Moe93} and has served as a
test of a variety of nuclear models.
It is the best available proof we can impose to a nuclear model
used to predict the $\beta\beta_{0\nu}$ matrix elements.

Many experimental groups have reported measurements of \bb ~
processes\cite{Moe93}. Nearly for all the cases the ground state (g.s.) to
ground
state ($0^+ \rightarrow 0^+$) decay was investigated. In direct-counting
experiments the analysis of the sum-energy spectrum of the emitted
electrons allowed the identification of the different $\beta\beta$-decay
modes.
The experimental research on the double
 beta decay of heavy deformed nuclei enjoys an increasing
activity. Thus, the \bb ~ half-life of  $^{238}U$\cite{Tur91}  and
$^{150}Nd$\cite{Moe93,Moe94,Art93}
have been measured and search is in progress for $^{244}Pu$\cite{Moo92}.

In previous papers \cite{Hir92,Cas93} we used the pseudo SU(3) shell model to
evaluate the
two neutrino double beta half lifes of eleven heavy deformed potential
double beta emitters. We found good agreement with the available
experimental information. The radiochemically measured $^{238}U$ decay
\cite{Tur91} raised some expectations, given the experimental arrangement
cannot
discriminate between the different double beta decay modes, and some
theoretical estimates\cite{Sta90} predicted similar decay ratios for both
the
$0\nu$ and the $2\nu$ modes. Our calculations for the two neutrino mode
were consistent with the experimental result.

We have also calculated the zero neutrino matrix elements for six
heavy deformed double beta emitters\cite{Hir94}.
 Using the upper limit for
the neutrino mass we also estimated their $\beta\beta_{0\nu}$ half-lives.
In the case of $^{238}U$
we found the zero neutrino half-life at least three orders of
magnitude greater than the two neutrino one, giving strong support of the
identification of the observed half-life as being the two
neutrino double beta decay.
The \bb ~of \nd ~to the ground and excited states of \sm ~has also
been studied\cite{Hir95}.
In the present paper we review the above mentioned calculations as
well as the pseudo SU(3) formalism used in those works.

\bigskip
\section{The pseudo SU(3) formalism}
\bigskip
\bigskip

In the pseudo $SU(3)$ shell model coupling scheme\cite{Rat73} normal
parity orbitals $(\eta ,l,j)$ are identified with orbitals of a harmonic
oscillator of one quanta
less $\tilde \eta = \eta-1$.  This set of orbitals with $\tilde j
= j = \tilde  l + \tilde s$,  pseudo spin $\tilde s =1/2$ and  pseudo
orbital angular momentum  $\tilde l$
define the so called pseudo space. The orbitals with
$j = \tilde l \pm 1/2$ are nearly degenerate. For configurations
of identical particles occupying a single j orbital of abnormal parity
a convenient characterization of states is made by means of the seniority
coupling scheme.

The many particle states of $n_\alpha$ nucleons in a given shell
$\eta_\alpha$,
$\alpha = \nu $ or $\pi$, can be defined by the totally antisymmetric
irreducible representations
$\{ 1^{n^N_\alpha}\} $ and $\{1^{n^A_\alpha}\}$ of unitary groups.
The dimensions of the normal $(N)$ parity space is
$\Omega^N_\alpha = (\tilde\eta_\alpha + 1) (\tilde\eta_\alpha +2)$ and
that of the unique $(A)$ space is $\Omega^A_\alpha =
2\eta_\alpha +4$ with the constraint
$n_\alpha = n^A_\alpha  + n^N_\alpha$.
Proton and neutron states are coupled to angular momentum $J^N$ and $J^A$
in both the normal and unique parity sectors, respectively. The
wave function of the many-particle state  with angular
momentum $J$ and projection $M$ is expressed as a direct product of the
normal  and unique parity ones, as:

\begin{equation}
|J M > = \sum\limits_{J^N J^A} [|J^N> \otimes |J^A>]^J_M
\end{equation}
We are interested in the low energy states which
have $J=0,2,3,4,6$. In the normal subspace
only pseudo spin zero configurations are taken into account.
Additionally in the abnormal parity space only seniority zero
configurations
 are taken into account.  This simplification implies that
 $J^A_\pi = J^A_\nu = 0$. This is a very
strong assumption quite useful in order to simplify the calculations.
Its effects upon the present calculation are discussed below.

The double beta decay, when described in the pseudo SU(3) scheme, is
strongly dependent on the occupation numbers for protons and neutrons in
the normal and abnormal parity states $n^N_\pi, n^N_\nu, n^A_\pi,
n^A_\nu$\cite{Cas93}.
These numbers are determined filling the Nilsson levels from below, as
discussed in \cite{Cas93}.
In particular the \bb ~decay is allowed only if
they fulfil the following relationships

\begin{equation}
\begin{array}{l}
n^A_{\pi ,f} = n^A_{\pi ,i} + 2~~,
\hspace{1cm}n^A_{\nu ,f} = n^A_{\nu ,i}~~, \\
n^N_{\pi ,f} = n^N_{\pi ,i}~~ ,
\hspace{1.7cm} n^N_{\nu ,f} = n^N_{\nu ,i} - 2 ~~.\label{num}
\end{array}
\end{equation}

\noindent
If they do not, the \bb ~decay becomes forbidden. This is the first selection
rule the pseudo SU(3) formalism impose to the double beta decay.
As an example, for \nd ~we have  obtained the occupation numbers
$n^A_\pi = 4,~n^N_\pi =6,~n^A_\nu = 2,~n^N_\nu = 6$.

We have
selected the standard version of the pseudo SU(3) Hamiltonian\cite{Dra84}.
It is constructed by a spherical Nilsson hamiltonian which describes the
single-particle motion of neutrons or protons, a quadrupole-quadrupole
interaction and a residual force.  The latter allows the fine tuning of
low lying spectral features like $K$ band splitting and the effective
moments of inertia.

With the occupation numbers  and the hamiltonian discussed above,
the wave function of the deformed ground state of \nd ~can be
written\cite{Cas93}

\begin{equation}
\begin{array}{ll}
|^{150}Nd, 0^+\rangle  = &
 | \ (h_{11/2})^4_\pi,  \ J^A_\pi = M^A_\pi = 0; \
(i_{13/2})^2_\nu ,\ J^A_\nu = M^A_\nu = 0 >_A \\
& | \ \{ 1^6\}_\pi  \{2^3\}_\pi (12,0)_\pi;
\ \{ 1^{6}\}_\nu \{2^3\}_\nu (18,0)_\nu; \ 1 (30,0) K=1 J = M = 0 >_N
\ \ ,  \label{nd}
\end{array}
\end{equation}

\bigskip
\section{\bf The double beta decay }
\bigskip

The inverse half life of the two neutrino mode of the $\beta\beta$-decay
can be expressed in the form\cite{Doi85}

\begin{equation}
	\left[\tau^{1/2}_{2\nu}(0^+ \rightarrow J^+_\sigma)\right]^{-1} =
      G_{2\nu}(J^+_\sigma) \ | \ M_{2\nu}(J^+_\sigma) \ |^2 \ \ .
\end{equation}

\noindent
where $G_{2\nu}(J^+_\sigma)$ are kinematical factors.
They depend on $E_{J;\sigma} = {1 \over 2} [{\it Q}_{\beta  \beta}-
E(J_\sigma )] + m_e c^2$ which is the half of total energy released. The
nuclear matrix element is evaluated using the pseudo SU(3) formalism.
For the \nd $\rightarrow$ \sm ~case it can be written as

\begin{equation}
\begin{array}{ll}
M_{2\nu}^{GT}(J^+_\sigma)~ =~a(J)~b(n^A_\pi)  ~{\cal E}_{J\sigma}^{-(J+1)}\\
\hspace{1cm}\sum\limits_{(\lambda_0 \mu_0 ) K_0}
<(0 \tilde\eta )1 \tilde l,(0 \tilde\eta )1 \tilde l \| (\lambda_0 \mu_0
) K_0 J>_1 \sum\limits_\rho
<(30,0)1~0,(\lambda_0 \mu_0 )K_0 J\|(\lambda \mu )_\sigma 1 J>_\rho\\
\hspace{1cm}\sum\limits_{\rho'}
\left[\begin{array}{cccc} (12,0) &(0,0) &(12,0) &1\\
(18,0) &(\lambda_0 \mu_0 ) &(12,2) &\rho' \\
(30,0) &(\lambda_0 \mu_0 ) &(\lambda \mu )_\sigma &\rho \\
1 &1 &1 \end{array} \right]
<(12,2)\mid\mid\mid [\tilde a_{0 \tilde \eta),{1\over 2}}
\tilde a_{0 \tilde \eta),{1\over 2}}]^{(\lambda_0 \mu_0 )}
\mid\mid\mid (18,0)>_{\rho'} \label{mgt}
\end{array}
\end{equation}

In the above formula $<..,..\|,,>$ denotes the SU(3) Clebsch-Gordan
coefficients\cite{Dra73}, the symbol  $[...]$ represents a
$9-\lambda\mu$ recoupling coefficient\cite{Mil78},
$<..\mid\mid\mid ..\mid\mid\mid ..>$ is the triple reduced matrix
elements\cite{Hir95c} and

\begin{equation}
\begin{array}{l}
a(0) = {{4\eta}\over{(2\eta + 1)\sqrt{2\eta - 1}}},\hspace{.5cm}
a(2) = {2\over {2\eta + 1}}
\left[{ {5 \eta (\eta -1))2\eta -3)}\over{3(2\eta +1)}}\right]^{1/2},
\hspace{.5cm}
b(n^A_\pi ) = [(n^A_\pi + 2)(\eta +1 - n^A_\pi/2)]^{1/2} ,\\ ~\\
	    {\cal E}_{J\sigma} =
E_{J\sigma} -\hbar \omega k_\pi 2 j_\pi + \Delta_C .\hspace{.5cm}
\Delta_C ={ 0.70 \over A^{1/3}} [2 Z + 1 - 0.76 ( (Z+1)^{4/3} -Z^{4/3} )]
  MeV~~.
\end{array}
\end{equation}
The SU(3) tensorial componentes $(\lambda_0 ,\mu_0)$ of the normal part
of the double Gamow-Teller operator must be able to couple the proton and
total irreps (18,0) and (30,0) associated with the ground state of \nd ~
to the corresponding irreps (12,2) and $(\lambda \mu)_\sigma )$ which
characterize the ground and excited rotational bands in \sm . If it is not
possible the \bb ~decay to the $0^+$ and $2^+$ states of an excited band is
forbidden. This is the
second selection rule imposed by the model to the \bb ~decay.

In the case the $0\nu$ decay exists, the virtual neutrino must be
emitted in
one vertex, and absorbed in the other. Since in the standard theory the
emitted particle is a right-handed antineutrino and the absorbed one a
left-handed neutrino the process requires that a) the exchanged neutrino
is a Majorana particle and b) both neutrinos have a common helicity
component.
The helicity matching can be satisfied in two ways: a) the neutrinos have
a nonvanishing mass and therefore a ``wrong'' helicity component
proportional to $m_\nu / E_\nu$. The decay rate will be proportional to
$<m_\nu >^2$. Or b) the helicity restriction could be satisfied if there
is a right handed current interaction. In this case a nonvanishing
mass allowing mixing of neutrino types is also required \cite{Doi85,Moe94b}.

The $\beta\beta_{0\nu}$ matrix elements were evaluated using the pseudo
SU(3) formalism\cite{Hir95}. The final expressions are similar to those
related with the two neutrino mode, given in Eq. (5), but restricted to the
decay to the ground state of the final nuclei. They explicitly include the
``neutrino potential'' originated by the interchange of a Majorana neutrino
between the two emitting neutrons. The $0\nu$ mode allows transitions  with
$\Delta l \neq 0$, whose contribution to the $\beta\beta_{0\nu}$
matrix elements were found an order of magnitude lesser than those which do
not change the orbital angular momentum.

The five nuclei which were found to have the
$\beta\beta_{2\nu}$ mode strictly forbidden within the pseudo SU(3)
model are allowed to have  $\beta\beta_{0\nu}$ decays
because the neutrino potential allows
parity mixing transitions. We would expect that these $0\nu$ matrix
elements, related with $\Delta l = 1,3,5$ transitions were of the same
order of magnitude of the $\Delta l = 2,4,6$ transitions
and about a factor ten lesser than the $\Delta l =
0$ matrix elements. This hindrance is comparable with the modifications that
the symplectic model would introduce, allowing more than one shell
active
for each kind of nucleons. These questions are under current investigation.

\section{Results and discussion}
\vskip 1.0pc

We predict, according to the assumptions of the model, that the two neutrino
double beta transitions of the following five nuclei:
$$
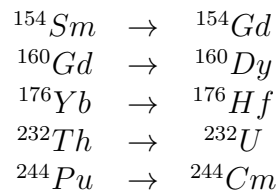
\matrix{^{154}Sm & \to & ^{154}Gd \cr
^{160}Gd & \to & ^{160}Dy \cr
^{176}Yb & \to & ^{176}Hf \cr
^{232}Th & \to & ^{232}U \cr
^{244}Pu & \to & ^{244}Cm \cr}
$$
{\it are forbidden}.
In particular we predict a null result for the present search in
$^{244}Pu$\cite{Moo92}, a potential strong test of our model.
 Inclusion of pairing within the same truncation scheme will allow
mixing of different pseudo SU(3) irreps, and is in progress\cite{Tro95}.
The lack of a
spin-isospin channel in our Hamiltonian is perhaps an additional limitation,
but it is included, at least partially, in the quadrupole-quadrupole
interaction, when it is recoupled.

In Table 1 are indicated the results
for the six $ \beta\beta$ emitters in which we get a \bb ~matrix element
different from zero\cite{Cas93}. The value for the integrated kinematical
factors were obtained following the procedure indicated
 by Doi et al\cite{Doi88} with $g_A/g_V = 1.0$.
 In the second and third
columns the theoretical and experimental \bb -half lives are given.
The agreement with the available data
for $^{150}Nd$ and $^{232}U$ is good. Improvements in the experimental
information about the decay of these nuclei together with $^{148}Nd$ is
possible, but the other three are almost excluded of detection because their
lepton kinematical factors $G_{2 \nu}$ are very small. This is
particularly
 true for $^{146}Nd$, which has the smallest Q value for the $\beta\beta$
decay and thus becomes  strongly inhibited.

We also evaluated the $\beta\beta_{0\nu}$ matrix elements  between the
the same six  nuclei\cite{Hir94}. In the last two
columns of Table 1 the theoretical predictions  and experimental
lower limits of the $\beta\beta_{0\nu}$-half lives are given.
In order to calculate the zero-neutrino
half-life we assumed  $<m_\nu>=1eV$, which is
the larger neutrino mass parameter compatible with the $^{76}Ge$ experiment
\cite{Mai94}. In the case of $^{238}U$ the predicted $0\nu$ half
life is three orders of magnitude greater than the predicted $2\nu$ half
life, which essentially agrees
with the experimental one, confirming that the observed $\beta\beta$
decay of $^{238}U$ has to be the two neutrino mode.
The
transition is strongly dominated by a pair of neutrons in the normal
parity orbital $i_{11/2}^\nu$
decaying into two protons in the unique orbital $i_{13/2}^\pi$,

In the case of $^{150}Nd$, the pseudo SU(3) $0\nu$ matrix element
reported here is about a factor four lesser than the QRPA
estimations\cite{Sta90}.
This is a very relevant result. First, it exhibits the stability of the
neutrinoless double beta decay matrix elements evaluated in quite
different nuclear models, in the case of deformed nuclei. Second, this
factor of four, which is little compared with the order of magnitude
variations in the $2\nu$ theoretical estimations, is still important in
order to extract the parameter $<m_e>$.
In this case
the transition is  dominated by a pair of neutrons in the normal
parity orbital $h_{9/2}^\nu$
decaying into two protons in the unique orbital $h_{11/2}^\pi$, again
resembling  the two neutrino case.

As can be seen in the last two columns of Table 1, the
$\tau_{0\nu}^{1/2}$ predicted for $<m_\nu > = 1 eV$ are at least three
order of magnitude greater than the experimental limits. These results
reflect the fact that at, the present stage of the experimental
$\beta\beta$ research, the limits $<m_\nu > \le 1.1 eV$ obtained by the
Heidelberg-Moscow collaboration \cite{Mai94} using significative volumes
of ultrapure $^{76}Ge$ are the most sensitive. But, if the
$\beta\beta_{0\nu}$ decay is
observed in $^{76}Ge$, at least a second observation will be essential,
and $^{150}Nd$ is a likely candidate to do this job \cite{Moe94,Moe94b}.
In the next few years the limit for $<m_\nu >$ extracted from
$\beta\beta_{0\nu}$ experiments is expected to be improved up to 0.1 eV
and $^{150}Nd$ is one of the selected isotopes \cite{Nem94}.

Finally we review the two neutrino mode of the double beta decay
\bb ~of
 \nd ~into the ground state, the first excited $2^+$ and the first  and
second excited $0^+$ states of \sm\sm
 \cite{Hir95}.

In Table 2 the matrix elements and predicted  half-lives
for the \bb ~decay of \nd ~to
the ground state, the first $2^+$ and the first and second excited $0^+$
states of \sm ~are presented.
The matrix elements  are given in units of $(m_e c^2)^{-(J+1)}$.
It must be mentioned that the phase space
factors differ in about $10 \%$ with those presented in Table 1\cite{Cas93}
where a different renormalization procedure was used.

The \bb ~decay to the first excited $0^+$ state is {\em forbidden}.
In this model this is imposed by the fulfilment of an
exact selection rule.
The pair of annihilation operators $\tilde a_{(0
\tilde 4 ){1\over 2}}$, when expanded in their SU(3) components,
cannot couple the \nd ~g.s. irrep (30,0)  to the irrep
(20,4) which we associated with the first excited $0^+$ state. Thus
the transition between members of these particular irreps are forbidden.

The decay to the second excited $0^+$ state is allowed but strongly
cancelled. The predicted
half-life is four orders of magnitude larger than that of the decay to
the g.s.
The \bb ~decay to the $2^+$ state is inhibited by the $\mu_N^3$
dependence of the matrix element as it is discussed in Section 3.
The matrix element of the \bb decay to the first excited $2^+$
state $M_{2\nu}(2^+_{g.s.} ) $ is three orders of magnitude lesser than
the matrix element of the decay to the g.s. $M_{2\nu}(0^+_{g.s.} ) $.

The present results contradicts those previously
published\cite{Pie94,Bar90} were it was speculated that the \bb ~decay of
\nd ~to the first excited $0^+$ state of \sm ~could have a similar intensity
of that to the g.s. We found that in the present formalism  this decay
is forbidden. If we select different
occupation numbers for both \nd ~and \sm , taken the deformation of the
latter nucleus instead of that of the former, we found very similar
results for the decay to the g.s. and the first $2^+$ state, but the
matrix elements of the decay to the first and second $0^+$ states becomes
interchanged with essentially the same values. Considering the difference in
the phase space integrals we predict a half live of the order
$10^{21}$ years for the decay to the first excited $0^+$ state and the
decay to the second excited one becomes forbidden.

The above discussed reduction of the matrix element of the \bb ~decay to
the excited $0^+$ state as compared with the decay to the g.s. is not a
general result of the pseudo SU(3) scheme. A recent analysis of the case of
$^{100}Mo$\cite{Hir95b} shows that both matrix elements are very similar and
that they are
in agreement with the experimental information.
In conclusion, the appearance of selection rules which can produce the
suppression of the matrix elements governing a \bb ~transition is a
consequence of the details of the irreps involved.

\bigskip

\section{Conclusions}

The pseudo SU(3) model is a very powerful machinery to describe the
collective behavior of heavy deformed nuclei.
It has been used to reproduce very accurately the
rotational spectra of heavy deformed nuclei, including the K-band splitting,
the amplitudes for transitions of the E2, M1 and M3 type.

We have used this model to evaluate the $\beta\beta_{2\nu}$ half lives of 11
heavy deformed nuclei, using the pseudo SU(3) approach together with a
summation
method, which avoids the closure approximation, obtaining good agreement with
the available data and making testable predictions, quite different
from the QRPA ones in some cases.

We have also  evaluated the $\beta\beta_{0\nu}$ half lives of six of these
heavy deformed nuclei. We obtained an order of magnitude agreement with
the QRPA estimations, and a factor four of difference. We
exhibited predictions for the neutrinoless double beta half-lives
assuming $<m_\nu>=1 eV$, and discussed the relevance of our results.
In the case of $^{238}U$ this result complements those obtained for the two
neutrino $\beta\beta$ decay, and it confirms the observed half-life as two
neutrino in origin.

At last we have studied the \bb ~decay mode
of \nd ~to the ground and excited states of \sm .
The \bb ~decay to the first excited $0^+$ state was found forbidden in
the model and the decay to the second excited $0^+$ state has a half-life
four orders of magnitude greater than that to the g.s..
The decay to the $2^+$ state is strongly inhibited due to the energy
dependence of the matrix elements $M_{2\nu}(2^+)$,  two powers greater than
that of the matrix element $M_{2\nu}(0^+)$.

In all the calculations only one active shell was allowed for protons, and
one for neutrons.
This is a very strong truncation, of the same type as used in shell model
 calculations.  An important consequence of this truncation is the fact
that only one
uncorrelated Gamow-Teller transition is allowed: that which removes a
neutron from a normal parity state with maximum angular momentum, and
creates a proton in the intruder shell
($h^\nu_{9/2} \rightarrow h^\pi_{11/2}$ in rare earth nuclei,
$i^\nu_{11/2} \rightarrow i^\pi_{13/2}$ in actinides).
This unique Gamow-Teller transition controls the double beta decay. If
 the occupation of the Nilsson levels is such that the number of protons
 in the abnormal states does not change going from the initial to the
final state configurations, the decay becomes forbidden.

The pseudo SU(3) model uses a quite restrictive Hilbert space. The model
could be improved by incorporating mixing between different irreps, via
pairing by example\cite{Tro95}. Also other active shells can be taken
into account in the symplectic extension\cite{Cas92b}. In both cases the
selection rules that impose such strong restrictions to the \bb ~decays
of some nuclei can be superseded.
However if the main part of the wave function is well
represented by the pseudo SU(3) model those forbidden decays will have, in
the better case, matrix elements that will be no greater than $20\%$ of
the allowed ones, resulting in at least one order of magnitude cancellation in
the half-life.

\bigskip
\centerline{\bf{Acknowledgements:}}
\bigskip
This work was supported in part by CONACyT under contract 3513-E9310.
The articles reviewed here were done in collaboration with Octavio Casta\~nos,
Peter O. Hess and O. Civitarese.

\newpage
\centerline{{\bf  Table Captions}}
\vskip 1.0truecm
\noindent
{\bf Table 1.}  Theoretical estimates for the $\beta\beta$-decay half-life
 in the $2\nu$ and $0\nu$ mode for several heavy deformed
nuclei are given and compared with the available experimental
data.

\vskip .5cm
{\bf Table 2} The dimensionless matrix elements
and predicted half-lives  for the \bb ~decay of \nd ~to
the ground state, the first $2^+$ and the first and second excited $0^+$
states of \sm .
\vfill
\newpage

\centerline{{\bf Table 1}}
$$
\matrix{
\noalign{\hrule}
\noalign{\vskip .5truecm }
 &~~~~~~~~ \tau^{1/2}_{2\nu} & [yr]~~~~~~~~~~~~~~~~~~~~~~~~~~~~
 &~~~~~~~~ \tau^{1/2}_{0\nu} & [yr]~~~~~~~~~~~~~~~~~\cr
 \hbox{Transition} & theo & exp
 & theo & exp \cr
\noalign{\vskip .5truecm}
\noalign{\hrule}
\noalign{\vskip .5truecm}
^{146}Nd \to ^{146}Sm & 2.1 \times 10^{31} &&1.18 \times 10^{28} &  \cr
\noalign{\vskip .5truecm}
^{148}Nd \to ^{148}Sm & 6.0 \times 10^{20} &&6.75 \times 10^{24} \ \cr
\noalign{\vskip .5truecm}
^{150}Nd \to ^{150}Sm &6.0 \times 10^{18} & 9 - 17\times 10^{18} \,
\cite{Moe93,Moe94,Art93}
&1.05 \times 10^{24} &> 2.1 \times 10^{21} \, \cite{Moe94}\cr
\noalign{\vskip .5truecm}
^{186}W \to ^{186}Os & 6.1 \times 10^{24}  &
&5.13 \times 10^{25} &> 2.3 \times 10^{20} \, \cite{Dan93}\cr
\noalign{\vskip .5truecm}
^{192}Os \to ^{192}Pt & 9.0 \times 10^{25} &&3.28 \times 10^{26}\cr
\noalign{\vskip .5truecm}
^{238}U \to ^{238}Pu & 1.4 \times 10^{21} & 2 \times 10^{21} \,\cite{Tur91}
&1.03\times 10^{24} &> 2.0 \times 10^{21} \, \cite{Tur91}\cr
\noalign{\vskip .5truecm}\noalign{\hrule}
\cr}
$$

\vskip 1cm

\centerline{Table 2}
$$
\begin{array}{llcccc}
\hline
&\vline\\
&\vline &M_{2\nu}^{GT}(J^+_\sigma)
&\tau^{1/2}_{2\nu}(0^+ \rightarrow J^+_\sigma)[yr] \\  &\vline \\ \hline
&\vline \\
 0^+\rightarrow 0^+(g.s.) &\vline &.0549
&6.73\times 10^{18}\\~  &\vline \\
0^+\rightarrow 0^+(1)&\vline &0
&\infty \\~  &\vline \\
0^+\rightarrow 0^+(2)&\vline &.00499
&4.31\times 10^{22} \\~  &\vline \\
 0^+\rightarrow 2^+ &\vline &5.38 \times 10^{-5}
&7.21 \times 10^{24}\\ ~  &\vline \\ \hline  \\
\end{array}
$$

\end{document}